\documentclass[11pt,a4paper]{article}
\pdfoutput=1
\usepackage[english]{babel}
\usepackage[latin1]{inputenc}
\usepackage{amsfonts,amsbsy,bm,euscript,mathrsfs}
\usepackage{amssymb,stmaryrd,faktor,slashed}
\usepackage[x11names]{xcolor}
\usepackage[tbtags]{amsmath}
\usepackage[bookmarks=true,colorlinks=true,linkcolor=black,citecolor=black,urlcolor=black,bookmarksnumbered]{hyperref}
\usepackage[nosort]{cite}

\numberwithin{equation}{section}

\makeatletter
\renewcommand\section{\@startsection {section}{1}{\z@}
{-3.5ex \@plus -1ex \@minus -.2ex}
{2.3ex \@plus.2ex}
{\normalfont\Large\bfseries}}
\renewcommand\subsection{\@startsection{subsection}{2}{\z@}
{-3.25ex\@plus -1ex \@minus -.2ex}
{1.5ex \@plus.2ex}
{\normalfont\large\bfseries}}
\makeatother

\newcommand{\foot}[1]{\footnote{#1\vspace{2pt}}}

\DeclareMathOperator{\arcsinh}{arcsinh}

\DeclareMathOperator{\arctanh}{arctanh}

\def\ads{{\rm AdS}_5\times {\rm S}^5}

\begin{document}

\setcounter{equation}{0}
\setcounter{footnote}{0}
\setcounter{section}{0}

\thispagestyle{empty}

\begin{flushright} \texttt{HU-EP-16/15\\ HU-MATH-16/10}\end{flushright}

\begin{center}
\vspace{1.5truecm}

{\LARGE \bf On jordanian deformations of AdS$_{\mathbf{5}}$ and supergravity}

\vspace{1.5truecm}

{Ben Hoare$^{1}$ and Stijn J. van Tongeren$^{2}$}

\vspace{1.0truecm}

{\em $^{1}$ Institut f\"ur Theoretische Physik, ETH Z\"urich,\\ Wolfgang-Pauli-Strasse 27, 8093 Z\"urich, Switzerland}

\vspace{0.5truecm}

{\em $^{2}$ Institut f\"ur Mathematik und Institut f\"ur Physik, Humboldt-Universit\"at zu Berlin, \\ IRIS Geb\"aude, Zum Grossen Windkanal 6, 12489 Berlin, Germany}

\vspace{1.0truecm}

{{\tt bhoare@ethz.ch, \quad svantongeren@physik.hu-berlin.de}}

\vspace{1.0truecm}
\end{center}

\begin{abstract}
We consider various homogeneous Yang-Baxter deformations of the $\ads$ superstring that can be obtained from the $\eta$-deformed superstring and related models by singular boosts. The jordanian deformations we obtain in this way behave similarly to the $\eta$-deformed model with regard to supergravity: T dualizing the classical sigma model it is possible to find corresponding solutions of supergravity, which, however, have dilatons that prevent T dualizing back. Hence the backgrounds of these jordanian deformations are not solutions of supergravity. Still, they do satisfy a set of recently found modified supergravity equations which implies that the corresponding sigma models are scale invariant. The abelian models that we obtain by singular boosts do directly correspond to solutions of supergravity. In addition to our main results we consider contraction limits of our main example, which do correspond to supergravity solutions.
\end{abstract}

\newpage

\setcounter{equation}{0}
\setcounter{footnote}{0}
\setcounter{section}{0}

\tableofcontents

\section{Introduction}\label{sec:intro}

Integrability plays an important role in furthering our understanding of the AdS/CFT correspondence \cite{Maldacena:1997re}.\foot{For reviews see e.g. \cite{Arutyunov:2009ga,Beisert:2010jr}} This motivates the search for less symmetric instances of this correspondence where integrability nevertheless persists. A famous example of such an instance is a string on the Lunin-Maldacena background \cite{Lunin:2005jy,Frolov:2005ty,Frolov:2005dj} dual to the real $\beta$ deformation of planar supersymmetric Yang-Mills theory (sYM). In recent years a number of new integrable deformations of the $\ads$ string have been found
\cite{Delduc:2013qra,Sfetsos:2013wia,Hollowood:2014qma,Demulder:2015lva}, falling in a class of so-called Yang-Baxter sigma models \cite{Klimcik:2002zj,Klimcik:2008eq}. In this paper we focus on the deformations of the $\ads$ supercoset sigma model \cite{Metsaev:1998it} as introduced in \cite{Delduc:2013qra}. The original form of this deformation is known as the $\eta$ deformation and corresponds to quantum deforming the symmetry algebra of the superstring \cite{Delduc:2013qra,Arutyunov:2013ega,Delduc:2014kha}. The $\eta$ deformation is based on a so-called non-split (inhomogeneous) $r$ matrix, but the construction can be generalized to homogeneous $r$ matrices \cite{Kawaguchi:2014qwa} which lead to Drinfeld twisted symmetry \cite{vanTongeren:2015uha}. The main questions surrounding these models are whether they correspond to real strings -- i.e. whether the associated backgrounds solve the supergravity equations of motion, making the models Weyl invariant -- and if so, what their AdS/CFT interpretation is.

In terms of non-split deformations, by explicitly constructing the deformed action to quadratic order in the fermions, it was found that the standard form of the $\eta$ deformation, despite having $\kappa$ symmetry \cite{Delduc:2013qra}, does not correspond to supergravity \cite{Arutyunov:2015qva}. However, T dualizing the model at the classical level in all six remaining isometry directions results in a background that is compatible with the supergravity equations of motion \cite{Hoare:2015wia}. This background has a dilaton with linear terms involving the isometry directions preventing us from T dualizing back in supergravity.\foot{These terms cancel out in the combination $e^\Phi F$ that appears in the quadratic fermionic terms of the classical Green-Schwarz action.} Still, the classical equivalence of the two theories means that the $\eta$-deformed string should be scale invariant, with its background satisfying a set of modified supergravity equations \cite{Arutyunov:2015mqj}. The scale invariance of the deformed model is also expected to follow from its $\kappa$ symmetry.\foot{We thank A. A. Tseytlin and L. Wulff for sharing related results that appeared shortly after the present paper \cite{Wulff:2016tju}. Let us also note that in the context of the pure spinor approach it was observed that classical BRST invariance (the analog of $\kappa$ symmetry) does not imply the supergravity equations of motion if the background possesses isometries \cite{Mikhailov:2012id,Mikhailov:2014qka}.}

Some of the homogeneous deformations do correspond to solutions of supergravity on the other hand. The Lunin-Maldacena background mentioned above can be viewed as one for instance \cite{Matsumoto:2014nra}. In fact, the known homogeneous deformations split in two classes -- abelian and jordanian -- and given numerous examples \cite{Matsumoto:2014nra,Matsumoto:2014gwa,Matsumoto:2015uja} it is clear that the abelian class corresponds to so-called TsT transformations \cite{vanTongeren:2015soa} which manifestly take strings to strings. For jordanian deformations the situation is less clear cut \cite{Kawaguchi:2014qwa,Kawaguchi:2014fca,Matsumoto:2014ubv,vanTongeren:2015soa,vanTongeren:2015uha}, but in at least one case the bosonic background \cite{Kawaguchi:2014fca,Matsumoto:2014ubv} of a jordanian deformation was completed to a supergravity solution. This background has not been compared directly with the fermions of the deformed supercoset model, however, so the string theory interpretation of jordanian deformations remains an open question. The homogeneous deformations that have a string theory interpretation are conjectured to generically correspond to Drinfeld twists of $\mathcal{N}=4$ sYM, leading to noncommutative field theory when deforming $\mathrm{AdS}_5$ \cite{vanTongeren:2015uha}.

In this paper we will investigate homogeneous deformations that are in an appropriate sense ``as close as possible'' to the $\eta$ deformation and other related inhomogeneous deformations considered in \cite{Delduc:2014kha} and \cite{Hoare:2016ibq}. It turns out that we can obtain a number of jordanian and abelian deformations by infinitely boosting the $\eta$-deformed string, and thereby investigate their supergravity properties. The upshot of this analysis is that jordanian deformations (that can be obtained this way) do not correspond to supergravity, in the same way that the $\eta$-deformed string does not: upon appropriately T dualizing at the classical level there exists a supergravity solution that, however, has a dilaton preventing us from T dualizing back in supergravity. Like the $\eta$ model \cite{Arutyunov:2015mqj}, these models have $\kappa$ symmetry and solve the modified supergravity equations of \cite{Arutyunov:2015mqj} and hence should be scale invariant. This class of models in particular includes the jordanian deformation of \cite{Kawaguchi:2014fca,Matsumoto:2014ubv}, which means that the supergravity solution proposed there is not the one corresponding to the deformed coset model. The abelian deformations we can obtain from the inhomogeneous models scale differently under the boost and thereby result in supergravity solutions, in line with their interpretation as TsT transformations. In particular the $\eta$ model can be boosted to the gravity dual of noncommutative sYM \cite{Hashimoto:1999ut,Maldacena:1999mh} in agreement with previous observations \cite{Arutyunov:2015qva}. The full set of $r$ matrices and models that can be obtained in this way consists of nine distinct jordanian and four distinct abelian ones, up to automorphisms.

In addition to this main result, for completeness with regard to our previous paper \cite{Hoare:2016ibq}, in which we discussed contraction limits of the various inhomogeneous models, we also briefly consider contraction limits of our main model in the spirit of \cite{Pachol:2015mfa}. As suggested in \cite{Pachol:2015mfa} the first of these gives a plane wave, matching a direct deformation of flat space considered in a lower dimensional setting in \cite{Borowiec:2015wua}. The second of these gives a T dual of flat space, a natural analogue of the results of \cite{Pachol:2015mfa,Hoare:2016ibq}. The contractions of our jordanian model do correspond to supergravity, in contrast to the standard contraction of the $\eta$-deformed string \cite{Arutyunov:2015qva}.

In the next section we introduce the general class of deformed models we will be working with, including the associated $R$ operators and $r$ matrices. In section \ref{sec:exampleintro} we introduce our main jordanian example as the ``sum'' of two inhomogeneous deformations. We then discuss its link to the $\eta$-deformed model via an infinite boost in section \ref{sec:boostingmainexample}, using this to provide a supergravity solution for its T dual. In section \ref{sec:boostingfurther} we discuss all deformations we can obtain in this way from the standard $\eta$-deformed string and other inhomogeneous deformations, with details presented in appendix \ref{app:generalboosts}. In appendix \ref{app:contractions} we discuss contractions of our main example of a jordanian deformation.

\paragraph{Note added:} while the present paper was in preparation reference \cite{Kyono:2016jqy} appeared on the arXiv, in which the authors computed R-R fluxes for various abelian $r$ matrices and also the jordanian $r$ matrix considered in \cite{Kawaguchi:2014fca}. Though approaching the problem from a different angle, this paper has some overlap regarding the jordanian deformation of \cite{Kawaguchi:2014fca} and the fact that it does not correspond to a supergravity background.

In this paper we have demonstrated this feature for a wide class of jordanian backgrounds. Additionally, via their relation to the $\eta$-deformed background, we answer some of the open questions in \cite{Kyono:2016jqy} with regards to the modified supergravity equations that are satisfied by jordanian deformations.

\section{The deformed superstring action}
\label{sec:modelintro}

The deformations of the $\mathrm{AdS}_5 \times \mathrm{S}^5$ superstring action we are interested in are of the form \cite{Delduc:2013qra}\foot{Here $T$ is the would-be effective string tension, $h$ is the world sheet metric, $\epsilon^{\tau\sigma}=1$, $A_\alpha = g^{-1} \partial_\alpha g$ with $g\in \mathrm{PSU}(2,2|4)$, $\mathrm{sTr}$ denotes the supertrace, and $d_\pm = \pm P_1 + \frac{2}{1-\eta^2} P_2 \mp P_3$ where the $P_i$ are the projectors onto the $i$th $\mathbb{Z}_4$ graded components of the semi-symmetric space $\mathrm{PSU}(2,2|4)/(\mathrm{SO}(4,1)\times \mathrm{SO}(5))$ (super $\mathrm{AdS}_5 \times \mathrm{S}^5$).}
\begin{equation}
\label{eq:defaction}
S = -\tfrac{T}{2} \int d\tau d\sigma \tfrac{1}{2}(\sqrt{h} h^{\alpha \beta} -\epsilon^{\alpha \beta}) \mathrm{sTr} (A_\alpha d_+ J_\beta)
\end{equation}
where $J=(1-\eta R_g \circ d_+)^{-1}(A)$ with $R_g(X)=g^{-1} R(g Xg^{-1}) g$.  The operator $R$ is a linear map from $\mathfrak{g}=\mathfrak{psu}(2,2|4)$ to itself. Setting $\eta=0$ ($R=0$) gives the undeformed $\mathrm{AdS}_5 \times \mathrm{S}^5$ superstring action of \cite{Metsaev:1998it}. Now, provided $R$ is antisymmetric,
\begin{equation}
\mathrm{sTr}(R(m) n) = -\mathrm{sTr}(m R(n)),
\end{equation}
and satisfies the non-split modified classical Yang-Baxter equation (mcYBe)
\begin{equation}
\label{eq:mcYBe}
[R(m),R(n)] - R([R(m),n] + [m,R(n)])=[m,n],
\end{equation}
this deformed model is classically integrable and has a form of $\kappa$ symmetry \cite{Delduc:2013qra}. Considered over a complex semisimple Lie algebra the standard solution of the mcYBe is the operator multiplying positive roots by $i$, negative roots by $-i$, and the Cartan generators by $0$. This solution preserves the real form $\mathfrak{psu}(2,2|4)$ and so can be used there as well.

This deformation method can be extended to include solutions of the (homogeneous) classical Yang-Baxter equation (cYBe) \cite{Kawaguchi:2014qwa}, where the cYBe reads
\begin{equation}
\label{eq:cYBe}
[R(m),R(n)] - R([R(m),n] + [m,R(n)])=0.
\end{equation}
A simple way to see this is to rescale the $R$ operator above as $R = \alpha/(2\eta) \hat{R}$ and consider the limit $\eta\rightarrow 0$. This turns the mcYBe for $R$ into the cYBe for $\hat{R}$, and leaves a deformed action based on $\hat{R}$ where $2\eta$ is replaced by $\alpha$, except in $d_\pm$ where $\eta$ is set to zero, in line with the action as presented in \cite{Kawaguchi:2014qwa}. We will henceforth drop the hat on $\hat{R}$. This limit of the mcYBe to the cYBe will come back repeatedly below. The classical Yang-Baxter equation has two known classes of solutions: abelian and jordanian ones, which in particular cases can be combined and extended.

By using the Killing form of our semi-simple superalgebra (the supertrace) we can conveniently represent $R$ operators by $r$ matrices, i.e.
\begin{equation}
R(m) = \mathrm{sTr}_2(r (1\otimes m))
\end{equation}
with
\begin{equation}
r = \sum \alpha_{ij} t^i \wedge t^j \equiv \sum \alpha_{ij} (t^i \otimes t^j - t^j \otimes t^i) \in \mathfrak{g} \otimes \mathfrak{g},
\end{equation}
where the $t^i$ generate $\mathfrak{g}$, $\alpha_{ij} \in \mathbb{R}$, and $\mbox{sTr}_2$ denotes the supertrace over the second space in the tensor product. We will refer to both the operator $R$ and its matrix representation $r$ as the $r$ matrix, where the latter satisfies the (m)cYBe in the form
\begin{equation}
[r_{12},r_{13}]+[r_{12},r_{23}]+[r_{13},r_{23}] = \begin{cases} \Omega & \mbox{non-split},\\
0 & \mbox{homogeneous}.\end{cases}
\end{equation}
Here $r_{mn}$ denotes the matrix realization of $r$ acting in spaces $m$ and $n$ in a tensor product, and $\Omega$ is the appropriately normalized canonical invariant element of $\Lambda^3_\mathfrak{g}$. In this notation, the standard solution of the non-split mcYBe over the complexified algebra $\mathfrak{g}_{\mathbb{C}}$ is given by
\begin{equation}
\label{eq:standardnon-splitr}
r = i e_j \wedge f^j,
\end{equation}
where the $e_j$ and $f_j$ denote positive and negative roots of $\mathfrak{g}_\mathbb{C}$ respectively, and we sum over all of them. A jordanian solution of the cYBe is built using a pair of generators where one is a raising (or lowering) operator with respect to the other. The prototypical example of this is a combination of a positive (or negative) root and a Cartan generator
\begin{equation}
r = h \wedge e
\end{equation}
with $[h,e]=e$. Extended jordanian solutions have terms added to the basic $h \wedge e$ piece. The remaining case is that of abelian solutions of the cYBe, which are built on commuting generators, namely
\begin{equation}
r = a \wedge b
\end{equation}
with $[a,b]=0$. Fixing a real form will generically result in multiple inequivalent solutions, we will see examples of this below.

From the form of the action it is clear that the undeformed symmetries preserved by a Yang-Baxter deformation are generated by \cite{vanTongeren:2015soa}
\begin{equation}
\{ t \in \mathfrak{g} \, | \, R([t,x]) = [t,R(x)] \ \forall \, x \in \mathfrak{g} \},
\end{equation}
which are the generators that `commute' with the $r$ matrix. Beyond these, the $\mathfrak{psu}(2,2|4)$ symmetry of the sigma model is deformed as described above.

Given an $r$ matrix and a coset parametrization for $g$ -- i.e. a pair $(r,g)$ -- an explicit background for the sigma model follows by inverting the operator $1-\eta R \circ d_+$, and comparing to the standard (Green-Schwarz) superstring action. We will use deformation parameters $\varkappa =  2 \eta/(1-\eta^2)$ in the non-split case, and $\alpha$ in homogeneous cases.

\section{An extended jordanian deformation of Poincar\'e $\mathrm{AdS}_5$}
\label{sec:exampleintro}

In our previous paper \cite{Hoare:2016ibq} we investigated deformations of the $\mathrm{AdS}_5 \times \mathrm{S}^5$ superstring based on split and non-split solutions of the modified classical Yang-Baxter equation.\foot{Split solutions have an opposite sign for the inhomogeneous term of the mcYBe. Such solutions exist for $\mathfrak{su}(2,2)$  but not for $\mathfrak{su}(4)$ and hence not for the full superalgebra $\mathfrak{psu}(2,2|4)$.} There and here, we are mainly interested in deformations of $\mathrm{AdS}_5$, hence $\mathfrak{su}(2,2)$ at the algebraic level. In our conventions, we have the following three $\mathfrak{su}(2,2)$ non-split $r$ matrices
\begin{align}
\nonumber
r^{ns}_{0} & = -(m^{0}{}_{i}\wedge m^{i5}+ m^{1}{}_{3}\wedge m^{32}+m^{1}{}_{4}\wedge m^{42}) \ , \\
\nonumber
r^{ns}_{1} & = -(m^{1}{}_{i} \wedge m^{i2} + m^{0}{}_{3} \wedge m^{35} + m^{0}{}_{4} \wedge m^{45}) \ , \\
r^{ns}_{2} & = -(m^{3}{}_{i} \wedge m^{i4} + m^{1}{}_{0} \wedge m^{02} + m^{1}{}_{5} \wedge m^{52}) \ .
\label{eq:non-splitrmatrices}
\end{align}
and the split $r$ matrix
\begin{equation}
\label{eq:splitrmatrix}
r^s = m^{1}{}_{i}\wedge m^{i5} + m^{0}{}_{3}\wedge m^{32} + m^{0}{}_{4}\wedge m^{42},
\end{equation}
where the $m^{ij}$, $i,j=0,\ldots,5$ are the generators of $\mathfrak{so}(2,4)\simeq \mathfrak{su}(2,2)$, with $0$ and $5$ the negative signature directions. In terms of a concrete basis we work with \cite{Arutyunov:2009ga,Pachol:2015mfa}
\begin{equation}\label{mgen}
m^{ij} = \frac14 [\gamma^i,\gamma^j] \ , \qquad   m^{i5} = - m^{5i} = \frac12 \gamma^i \ , \qquad i = 0,\ldots,4 \ ,
\end{equation}
where
\begin{align}
\nonumber
\gamma^0 & = i \sigma_3 \otimes \sigma_0 \ , & \gamma^1 & = \sigma_2 \otimes \sigma_2 \ , & \gamma^2 & = - \sigma_2 \otimes \sigma_1 \ ,
\\
\gamma^3 & = \sigma_1 \otimes \sigma_0 \ , & \gamma^4 & = \sigma_2 \otimes \sigma_3 \ , & \gamma^5 & = - i \gamma^0 \ ,
\end{align}
$\sigma_0 = \mathbf{1}_{2\times 2}$ and $\sigma_a$ are the Pauli matrices. These $m^{ij}$ satisfy the standard $\mathfrak{so}(2,4)$ commutation relations
\begin{equation}
[m^{ij},m^{kl}] = \eta^{jk}m^{il} - \eta^{ik}m^{jl} - \eta^{jl} m^{ik} + \eta^{il}m^{jk} \ , \qquad i,j,k,l = 0,\ldots, 5 \ ,
\end{equation}
where $\eta = \operatorname{diag}(-1,1,1,1,1,-1)$. With respect to the projection operators in the action these generators are grade two when they have an index $5$ and grade zero otherwise.

In this paper we would like to investigate deformations based on solutions of the homogeneous classical Yang-Baxter equation that are as close as possible to solutions of the inhomogeneous (modified) equation in a well defined sense. The primary example of such an $r$ matrix is
\begin{equation}
r^j_0 = r^{ns}_{0} + r^s,
\end{equation}
which solves the cYBe despite the equation being quadratic.\foot{One reason to try this is that the $r$ matrices of the $\kappa$-Poincar\'e algebra \cite{Lukierski:1991pn} have this structure, see e.g. \cite{Borowiec:2013lca}.}

As will become clear below, the $r$ matrices we will consider naturally deform the Poincar\'e patch of anti-de Sitter space. Let us define the corresponding physical generators as
\begin{equation}
\begin{aligned}
\label{eq:physicalgenerators}
p^\mu & = m^{\hat{\mu}0}-m^{\hat{\mu}1}, \hspace{30pt} M^{\mu\nu} = m^{\hat{\mu}\hat{\nu}}, \\
k^\mu & = m^{\hat{\mu}0}+m^{\hat{\mu}1}, \hspace{30pt} D=-m^{01},
\end{aligned}
\end{equation}
where\foot{To fit established conventions we take the coset direction to be $5$, while here it is most convenient to take this to be the timelike direction in the Poincar\'e patch. The shift of the spatial indices arises because in our conventions it is natural to analytically continue the standard non-split $r$ matrix to the split one by either $0\rightarrow1$ or $0\rightarrow2$ as opposed to $0 \rightarrow 4$.}
\begin{equation}
\hat{\mu} = \begin{cases} 5 & \mu = 0, \\ i+1 & \mu = i \in \{ 1,2,3\}.\end{cases}
\end{equation}
The distinction between these indices will be clear from the context. In terms of these generators, we have
\begin{equation}
\label{eq:rh0}
r^j_0 = D \wedge p^0 + M^{0}{}_{\mu} \wedge p^\mu - M^{1}{}_2\wedge p^2 - M^{1}{}_3 \wedge p^3,
\end{equation}
which manifests the extended jordanian nature of $r^j_0$. This $r$ matrix preserves the $\mathbb{R}^2 \times \mathrm{U}(1)$ symmetry generated by $\{p^0, p^1, M^{23}\}$. Therefore the $\mathbb{R}^2$ generators correspond to null directions in the $\mathfrak{so}(2,4)$ algebra.

To find the deformation of $\mathrm{AdS}_5$ generated by $r^j_0$, we introduce the coset parametrization
\begin{equation}
\label{eq:ghinphysgen}
g^h = e^{-(t+z) p^0 + x p^1} e^{\theta M^{23}} e^{r p^2} e^{-\frac{1}{2z} k^0},
\end{equation}
where $z$ is the radial direction in AdS, and $t$, $x$ are Cartesian and $r$ and $\theta$ polar coordinates on the boundary Minkowski space, or equivalently
\begin{equation}
g = e^{x_\mu p^\mu} e^{-z p^0} e^{-\frac{1}{2z} k^0}.
\end{equation}
We have not previously encountered this type of coset parametrization for Poincar\'e AdS in the literature
.\foot{It is possible to use the coset direction $5$ instead of index $0$ in $p$ and $k$ to build up a coset representative of this type as well. This, however, apparently requires introducing $z$ in a less elegant fashion, giving a group element of the form
\begin{equation*}
g= e^{x_\mu (m^{\mu5}-m^{\mu4})}e^{z\sqrt{1+1/z}(m^{05}-m^{04})}e^{-\sqrt{1+1/z}(m^{05}+m^{04})}.
\end{equation*}}
Inserting the combination $(r^{j}_0,g^h)$ in the action gives
\begin{equation}
\begin{aligned}
\label{eq:hombackground0}
ds^2_0 & = \frac{dz^2-dt^2}{z^2 - \alpha^2} + \frac{dx^2 + dr^2}{z^2+ \alpha^2 r^2/z^2} +\frac{r^2 d\theta^2}{z^2},\\
B_0 & = -\frac{\alpha}{z}\frac{1}{z^2-\alpha^2} dt \wedge dz - \frac{\alpha r}{z^4+\alpha^2 r^2} dx \wedge dr.
\end{aligned}
\end{equation}
Note that we can set the deformation parameter to $1$ in this background by rescaling $x,r,t$ and $z$ by $\alpha$. This is a consequence of the automorphism of the conformal algebra given by $p^\mu \rightarrow a p^\mu$, $k^\nu \rightarrow a^{-1} k^\nu$. Since $r^j_0$ is homogeneous in $p$ and does not depend on $k$, by this automorphism a deformation based on $\alpha r^j_0$ is equivalent to one based on $r^j_0$ at the algebraic level, and the rescaling of coordinates simply reflects this, cf. eqn. \eqref{eq:ghinphysgen}. The same considerations apply to various backgrounds considered in e.g. \cite{vanTongeren:2015uha} and all other backgrounds we will consider below.

Completing the above deformation of $\mathrm{AdS}_5$ with an undeformed five sphere gives us the bosonic sector of our homogeneous deformed sigma model. In order to unambiguously determine the fermionic sector we should in principle work with the complete deformed supercoset sigma model. At the same time, there is a conjecture that all homogeneous Yang-Baxter deformations of $\mathrm{AdS}_5 \times \mathrm{S}^5$ correspond to solutions of supergravity -- this is sometimes referred to as the ``gravity/cYBe correspondence'' \cite{Matsumoto:2014nra,Matsumoto:2014cja,Matsumoto:2016lnr}. From this point of view, we may try to avoid the technical complications brought in by the fermions and simply try to find a solution of supergravity that incorporates our bosonic background, and at least has the manifest symmetries preserved by the deformation. This then provides a \emph{candidate} background for the deformed sigma model. As it turns out we can do better: this model based on $r^j_0$ is one of many models that can be obtained via infinite boosts of the standard non-split deformed supercoset model, whose fermions have been investigated in detail in \cite{Arutyunov:2015qva}. The resulting models split into two classes: jordanian (including the above one) whose fermions behave similarly to the fermions of the non-split deformation \cite{Arutyunov:2015qva,Hoare:2015wia,Arutyunov:2015mqj} and are incompatible with the supergravity equations of motion, and abelian whose fermions are compatible with supergravity as expected by their relation to TsT transformations \cite{vanTongeren:2015soa}. We will thus find a number of counterexamples to the ``gravity/cYBe correspondence'' conjecture of \cite{Matsumoto:2014nra,Matsumoto:2014cja,Matsumoto:2016lnr}.

\section{A boost of the standard non-split deformation}
\label{sec:boostingmainexample}

We can view $r^j_0$ as the result of an infinite boost applied to $r^{ns}_0$, where by a boost we simply mean a transformation generated by a noncompact generator of $\mathfrak{so}(2,4)$, not necessarily a Lorentz boost in terms of the physical generators. Because our Poincar\'e parametrization singles out indices $0$ and $1$, we would like to make sure any boost is compatible with this structure. We therefore introduce
\begin{equation}
b_{ij}=e^{\frac{\pi}{2}(m_{i0}-m_{j1})} e^{\beta m_{ij}},
\end{equation}
which rotates in addition to boosting by $\beta$ in the $ij$-plane, so that the result will have the desired index structure. For a large $\beta$ boost in the $01$-plane we then have\foot{Boosting oppositely gives $r^j_0$ with momenta replaced by special conformal generators up to signs. Given the automorphism of the conformal algebra under $D \rightarrow -D$ and $p\leftrightarrow k$ we consider this $r$ matrix equivalent to $r^j_0$ for our present purposes. The same considerations apply to the other examples below.}
\begin{equation}
\mbox{Ad}_{b_{01}}(r^{ns}_{0}) \sim \frac{e^{\beta}}{2} r^j_0.
\end{equation}
As only the combination of the deformation parameter and the $r$ matrix enters the action, by simultaneously scaling the deformation parameter $\varkappa \sim e^{-\beta}$ -- equivalently $\eta \sim e^{-\beta}$ -- we can obtain a finite result in the limit $\beta \rightarrow \infty$, i.e. with $\varkappa = 2 \alpha e^{-\beta}$
\begin{equation}
\label{eq:mainboostedrmat}
\lim_{\beta \rightarrow \infty} \mbox{Ad}_{b_{01}}(\varkappa r^{ns}_{0}) = \alpha r^j_0.
\end{equation}
The complete non-split deformation of $\mathrm{AdS}_5 \times \mathrm{S}^5$ including fermions is based on the extension of $r^{ns}_0$ to $\mathfrak{psu}(2,2|4)$. However, its fermionic and $\mathfrak{su}(4)$ terms are boost invariant and effectively vanish in this scaling limit, and hence the full $\mathfrak{psu}(2,2|4)$ $r^{ns}_0$ gives $r^j_0$.\foot{Moreover the $d_\pm$ operators in the action reduce to the ones appropriate for the homogeneous deformation, cf. the discussion of section \ref{sec:modelintro}.} We will come back to other the possible $r$ matrices we can obtain in this way later. Let us first understand what an infinite boost means in terms of the parametrization of our (coset) space, starting by recalling the parametrization and background of the standard non-split case.

\subsection{The T dual of the standard non-split deformation}

The original non-split deformation of the $\mathrm{AdS}_5 \times \mathrm{S}^5$ superstring sigma model is based on the standard $r$ matrix of eqn. \eqref{eq:standardnon-splitr} (the $\mathfrak{psu}(2,2|4)$ extension of $r^{ns}_0$) where in our conventions the $e_i$ are the strictly upper triangular matrix unities of the $4|4 \times 4|4$ supermatrix realization of $\mathfrak{su}(2,2|4)$ and $f^i = e_i^t$.\foot{Our conventions for $\mathfrak{su}(2,2)$ are specified above. These agree with the conventions of \cite{Arutyunov:2009ga}, which we follow for the full $\mathfrak{psu}(2,2|4)$.} The associated sigma model was worked out to second order in the fermions in \cite{Arutyunov:2015qva}, based on a coset parametrization of the form
\begin{equation}
g = g_0 g_s g_f,
\end{equation}
where
\begin{equation}
\label{eq:gns0}
g_0 = \lambda e^{\arcsin{x} \, m^{13}} e^{\arcsinh{\rho}\, m^{15}} \,
\end{equation}
with
\begin{equation}
\lambda =  e^{t m^{05} - \psi_1 m^{12} - \psi_2 m^{34}} \ .
\end{equation}
The remaining $g_s$ and $g_f$ parametrize the sphere and the fermions respectively, the details of which do not affect our present considerations. Formally T dualizing the model in all six remaining isometry directions including time, gives a standard Green-Schwarz action in the background\foot{Regarding supergravity we follow the conventions of \cite{Arutyunov:2015mqj}.} \cite{Hoare:2015wia}
\begin{equation}
\label{eq:non-splitTdualbackground}
\begin{aligned}
ds^2 = &  - \frac{1-\varkappa^2 \rho^2}{1+ \rho^2} d\hat{t}^2 + \frac{1}{1-\varkappa^2 \rho^2} \frac{d \rho^2}{1+\rho^2} + \frac{d \hat{\psi}_1^2}{\rho^2 (1-x^2)} + \frac{(\rho dx + \varkappa \rho x d\hat{\psi}_1)^2}{1-x^2} + \frac{d\hat{\psi}_2^2}{\rho^2 x^2} \\
& + \frac{1+\varkappa^2 \varrho^2}{1-\varrho^2} d\hat{\varphi}^2 + \frac{1}{1+ \varkappa^2 \varrho^2}\frac{d\varrho^2}{1- \varrho^2} + \frac{d\hat{\phi}_1^2}{\varrho^2 (1-w^2)} + \frac{(\varrho dw - \varkappa \varrho w \, d \hat{\phi}_1)^2}{1-w^2} + \frac{d \hat{\phi}_2^2}{\varrho^2 w^2}\\
B = & 0\\
e^\Phi F_5 = &  4 i \sqrt{1+\varkappa^2}(e^{\hat{t}} \wedge e^{\hat{\psi}_2} \wedge e^{\hat{\psi}_1} \wedge e^w \wedge e^\varrho) + \,\,\mbox{dual},
\end{aligned}
\end{equation}
where
\begin{equation}
\begin{aligned}
 e^{\hat{t}} & =\frac{1}{\sqrt{1+ \rho^2}}(d\hat{t} \, + \,\frac{\varkappa \rho}{1-\varkappa^2 \rho^2} d \rho)\,, \hspace{10pt} e^w = \frac{1}{{\sqrt{1-w^2}}}(\varrho\, dw - \varkappa \varrho w\, d \hat{\phi}_1)\,,\\
e^{\hat{\psi}_2} &  = \frac{ d\hat{\psi}_2}{\rho x}\,,\hspace{15pt} e^{\hat{\psi}_1} = \frac{d\hat{\psi}_1}{\rho \sqrt{1-x^2}}\,,\hspace{15pt}  e^\varrho = \frac{1}{\sqrt{1- \varrho^2}}\big( \frac{d\varrho}{1+\varkappa^2 \varrho^2} + \varkappa \varphi d\hat \varphi\big) \,.
\end{aligned}
\end{equation}
Here we use $\varrho$ instead of $r$ as used in \cite{Arutyunov:2013ega,Arutyunov:2015qva,Hoare:2015wia,Arutyunov:2015mqj}. This corresponds to a solution of type IIB$^*$ supergravity if we split off the dilaton as
\begin{equation}
\label{eq:etaTdualdilaton}
e^\Phi = e^{\Phi_0} e^{-4 \varkappa(\hat{t}+\hat{\varphi}) -2 \varkappa(\hat{\psi}_1-\hat{\phi}_1)}\frac{(1-\varkappa^2 \rho^2)^2(1+\varkappa^2 \varrho^2)^2}{4\rho^2 \varrho^2 \sqrt{1+\rho^2} \sqrt{1-\varrho^2} x \sqrt{1-x^2} w \sqrt{1-w^2}}.
\end{equation}
The dependence of the dilaton on the isometry directions shows that we cannot T dualize back at the level of supergravity -- i.e. beyond the classical level in the sigma model -- and hence that the background of the original deformed sigma model cannot solve the supergravity equations of motion. Based on the discussion above we expect that a certain $\varkappa\rightarrow0$ limit of this background should correspond to the T dual of our deformed model. We can determine the required scaling of the coordinates by comparing the group elements.

\subsection{Boosting from global to Poincar\'e AdS}

We would like take the group element of eqn. \eqref{eq:gns0}
\begin{equation}
g_0 =  e^{t m^{05} - \psi_1 m^{12} - \psi_2 m^{34}} e^{ \arcsin{x} \, m^{13}} e^{ \arcsinh{\rho}\, m^{15}},
\end{equation}
boost it an infinite amount, and if possible relate it to the parametrization of eqn. \eqref{eq:ghinphysgen}
\begin{equation}
g^h = e^{-(t+z) (m^{50} -m^{51}) + x (m^{20} - m^{21})} e^{\theta m^{34}} e^{r (m^{30} - m^{31})} e^{-\frac{1}{2z} (m^{50} + m^{51})},
\end{equation}
used in the homogeneous case. Keeping in mind that in the sigma model group elements are defined up to gauge transformations, which act via right multiplication by elements of the $\mathrm{SO}(4,1)$ generated by the $m^{ij}$ for $i,j\neq5$, we note the $z$ dependent part of the above group element
\begin{equation}
g^h_z = e^{-z (m^{50}-m^{51})} e^{-\frac{1}{2z} \, (m^{50}+m^{51})},
\end{equation}
is gauge equivalent to the $\rho$ dependent part of $g_0$, multiplied by the result of a constant shift of its $t$
\begin{equation}
g^h_z = e^{\tfrac{\pi}{2} m^{05}} e^{\arcsinh{\rho}\, m^{15}} b^{-1}_{\rho}
\end{equation}
with
\begin{equation}
b_\rho = e^{(\arcsinh{\rho}-\log 2) \, m^{01}},
\end{equation}
under the identification $z =\sqrt{1+\rho^2}-\rho$. The remaining terms in $g_0$ simply transform by $\mathrm{Ad}_{b_{01}}$ directly. This shows that up to gauge equivalence by $b b_\rho$ (where $b = b_{01}$) and a constant shift of $t$, boosting $g_0$ means
\begin{equation}
\begin{aligned}
&t\, m^{05} & \longrightarrow \hspace{20pt} &-\tfrac{1}{2}t\, (e^\beta p^0 +e^{-\beta} k^0),&\\
&\psi_1\, m^{12} & \longrightarrow \hspace{20pt} &\tfrac{1}{2}\psi_1\, (e^\beta p^1-e^{-\beta} k^1),& \\
&\arcsin{x}\,  m^{13} & \longrightarrow \hspace{20pt} &\tfrac{1}{2}\arcsin{x}\, (e^\beta p^2-e^{-\beta} k^2),& \\
&(\sqrt{1+\rho^2}\pm \rho) & \longrightarrow \hspace{20pt} & e^{\mp\beta} (\sqrt{1+\rho^2}\pm \rho).&
\end{aligned}
\end{equation}
If we now take $t = 2e^{-\beta} \tilde{t}$, $\psi_1= -2e^{-\beta} \tilde{x}$, $\psi_2 = -\theta$, $x= 2 e^{-\beta} \tilde{r}$ and $\rho =  e^\beta \frac{1}{2z}$, in the limit $\beta\rightarrow \infty$ this produces precisely the homogeneous parametrization $g^h$ from the global parametrization $g_0$ upon dropping tildes. At the level of the metric it is also clear that taking large $\rho$ gives Poincar\'e AdS from global AdS. Upon including the fermions we strictly speaking cannot use the above gauge equivalence anymore: we instead end up with $b b_\rho g_f$. Still we can do a gauge transformation by $(b b_\rho)^{-1}$ to give $b b_\rho g_f (b b_\rho)^{-1}$. This transformation can be absorbed in a redefinition of the fermions since $b b_\rho$ remains finite in the infinite boost limit. The coordinates on the sphere require no scaling.

\subsection{A jordanian T dual supergravity solution}

We can now implement this scaling in the full non-split background, where cf. eqn. \eqref{eq:mainboostedrmat} we take $\varkappa = 2 e^{-\beta} \alpha$. To include the fermions it is easiest to work with the T dualized background \eqref{eq:non-splitTdualbackground}. Here we should scale the T dual fields inversely, but otherwise the limit is the one discussed above. Dropping tildes, eqs. \eqref{eq:non-splitTdualbackground} become
\begin{equation}
\label{eq:rh0Tdualsugra}
\begin{aligned}
ds^2 = & \frac{dz^2}{z^2 - \alpha^2}-(z^2 -\alpha^2) d\hat{t}^2 + z^2 d\hat{x}^2 + \frac{(dr + \alpha r d\hat{x})^2}{z^2} + \frac{z^2}{r^2} d\theta^2\\
& + \frac{ d\hat{\varphi}^2 + d\varrho^2}{1-\varrho^2}  + \frac{d\hat{\phi}_1^2}{\varrho^2 (1-w^2)} + \frac{\varrho^2 dw^2}{1-w^2} + \frac{d \hat{\phi}_2^2}{\varrho^2 w^2}\\
B = & 0 \\
e^\Phi F_5 = &  4i \frac{\varrho}{r \sqrt{1-\varrho^2}\sqrt{1-w^2}}(z^3 d\hat{t} - \frac{\alpha z^2}{z^2 - \alpha^2} dz )\wedge d\hat{x} \wedge d \theta \wedge d \varrho \wedge d w + \,\,\mbox{dual}
\end{aligned}
\end{equation}
the metric and $B$ field of which are precisely T dual to eqs. \eqref{eq:hombackground0}.\foot{This follows directly by standard T duality in $t$, $x$, and $\theta$, if we drop the total derivative in the $B$ field. If this is taken along, as usual an additional coordinate redefinition in the T dual background is required.} Taking the same limit in the dilaton gives\footnote{Note that the identification of the dilaton from the $\sigma$ model as in eqs. (\ref{eq:non-splitTdualbackground}-\ref{eq:etaTdualdilaton}) is not unique; rescaling $e^\Phi$ by a constant and $F_5$ inversely, does not affect the supergravity equations or the $\sigma$ model Lagrangian. Here we take the $\eta$-model solution of eqs. (\ref{eq:non-splitTdualbackground}-\ref{eq:etaTdualdilaton}) and rescale $\Phi \rightarrow \Phi+2 \beta$ before the singular boost limit, to get a manifestly finite dilaton. Alternatively, our limit gives a $\sigma$ model with finite $e^\Phi F_5$, from which we extract the given dilaton.}
\begin{equation}
e^\Phi = e^{\Phi_0} e^{-4\alpha \hat{t}-2 \alpha \hat{x}} \frac{(z^2 - \alpha^2)^2}{2rz}\frac{1}{2\varrho^2 \sqrt{1-\varrho^2}w\sqrt{1-w^2}}.
\end{equation}
The dependence of the dilaton on the isometric directions $\hat{t}$ and $\hat{x}$ of the sigma model means that we cannot T dualize back at the level of supergravity, and therefore that the fermions of the $r^j_0$ deformed sigma model do not correspond to a solution of supergravity. However, as an immediate corollary of the method we have used to find this background, it does satisfy the modified supergravity equations of \cite{Arutyunov:2015mqj}, and hence the corresponding sigma model is scale invariant.

\section{Further boosts of the standard non-split deformation}
\label{sec:boostingfurther}

Given the invariance of the non-split $r$ matrices under pairwise permutations of indices $0$ and $5$, $1$ and $2$, and $3$ and $4$, there are two a priori inequivalent boosts we can consider. We can take these to be $b_{01}$, used above, and $b_{03}$. This alternative option turns out to make contact with the Hashimoto-Itzhaki-Maldacena-Russo (HIMR) background \cite{Hashimoto:1999ut,Maldacena:1999mh}. In addition to this we can consider doing two consecutive boosts, through which amongst others we can make contact with the jordanian deformation of \cite{Kawaguchi:2014fca}.

\subsection{The Hashimoto-Itzhaki-Maldacena-Russo background}

Boosting $r^{ns}_0$ with $b_{03}$ requires us to scale the deformation parameter differently from before. The leading term in the $r$ matrix now scales as $e^{2\beta}$ meaning we should scale $\varkappa \sim e^{-2\beta}$. Doing so gives
\begin{equation}
\lim_{\beta \rightarrow \infty} 4 \alpha e^{-2\beta} \mbox{Ad}_{b_{03}}(r^{ns}_0) = \alpha r^a_0,
\end{equation}
where
\begin{equation}
\label{eq:MRrmatrix}
r^a_0  = p^1 \wedge p^2
\end{equation}
is the abelian $r$ matrix that, as first observed in \cite{Matsumoto:2014gwa}, corresponds to the HIMR deformation of Poincar\'e AdS, the gravitational dual to canonical noncommutative SYM \cite{Hashimoto:1999ut,Maldacena:1999mh}.

To implement the infinite $b_{03}$ boost on the global group element $g_0$ we can simply rotate the discussion of the previous section in the $13$-plane by $\pi/2$ before doing the boost. In other words we start from $e^{\pi/2 m_{31}} g_0 e^{-\pi/2 m_{31}}$, on which $b_{03}$ acts exactly as $b_{01}$ did on $g_0$, giving $g^h$ under the same scalings as before. Now, the left action of $e^{\pi/2 m_{31}}$ on $g_0$ effectively interchanges $\psi_1$ and $\psi_2$ and replaces $x \rightarrow - \sqrt{1-x^2}$, while the right action by $e^{-\pi/2 m_{31}}$ is just a gauge transformation. Hence in a model based on $g_0$ we can implement the $b_{03}$ boost by the same scaling as before, up to expanding around $x=1$ and exchanging $\psi_1$ and $\psi_2$.  Implementing this limit on the non-split background based on $(r^{ns}_0,g_0)$ then gives the HIMR background (in polar coordinates), as expected from the $r$ matrix picture. A closely related limit of the non-split background to the HIMR one was previously found by S. Frolov \cite{Arutyunov:2015qva} -- our results simply explain this at the level of the coset construction.

In this boost limit the deformation parameter scales to zero twice as fast as the isometry coordinates. This is in contrast to the jordanian boost limit discussed in section \ref{sec:boostingmainexample} for which the deformation parameter scaled to zero at the same speed as the isometry coordinates. Consequently, in this case the problematic linear terms in the T dual dilaton disappear and hence the non-split background limits to a solution of supergravity, consistent with the TsT interpretation of the abelian $r$ matrix.

\subsection{Consecutive boosts}
\label{sec:consecutiveboosts}

Looking at the expressions for the $r$ matrices \eqref{eq:rh0} and \eqref{eq:MRrmatrix}, we can additionally Lorentz boost them to obtain light cone versions of these $r$ matrices. The possible Lorentz boosts are
\begin{equation}
B_{0j} = e^{-\pi/2 M_{j1}} e^{\beta M_{0j}},
\end{equation}
for $j=1,2,3$, where we again rotate the result for convenience. Boosting $r^j_0$ by $B_{01}$ results in further jordanian $r$ matrices
\begin{equation}
\lim_{\beta \rightarrow \pm \infty} \sqrt{2} e^{-|\beta|} \gamma \mbox{Ad}_{B_{01}} r^j_0 =  \gamma r^j_\pm
\end{equation}
where
\begin{equation}
\begin{aligned}
r^j_+ & = (D + M^{-}{}_-)\wedge p^- + 2 M^{-}{}_2\wedge p^2 + 2 M^{-}{}_3\wedge p^3,\\
r^j_- & = (D + M^{+}{}_+)\wedge p^+,
\end{aligned}
\end{equation}
and we have introduced light cone coordinates $\sqrt{2} a^\pm =a^0\pm a^1$. Using $M_{02}$ or equivalently $M_{03}$ requires a different scaling and produces the abelian $r$ matrix
\begin{equation}
\lim_{\beta \rightarrow \infty}  2e^{-2\beta} \gamma \mbox{Ad}_{B_{03}} r^j_0 = \gamma M^{3-} \wedge p^- \equiv \gamma r^a_1.
\end{equation}
Finally, we can boost the abelian $r$ matrix \eqref{eq:MRrmatrix} to $p^\pm \wedge p^2 \equiv r^a_\mp $.

In terms of the associated backgrounds, it is physically clear that the boost amounts to rescaling the corresponding light cone coordinates.\foot{In terms of the group element $g^h$ this immediately follows for everything but its $z$ dependent part. We have not attempted to find the gauge transformation that would manifest it also for this part. Its existence is ensured however, as $P_2(A^h) = P_2(\tilde{A}^h)$, where $A^h$ is the current built from $g^h$ and $\tilde{A}^h$ is one built from $B_{0j} g^h$.} Rather than writing down further explicit T dual supergravity backgrounds, let us briefly comment on the resulting models.

The deformation associated to $r^j_-$ has been previously considered in the literature \cite{Kawaguchi:2014fca,Matsumoto:2014ubv,vanTongeren:2015uha}. In particular the authors of\cite{Kawaguchi:2014fca}  conjectured a supergravity solution for this model. Now, implementing the appropriate limit in eqs. \eqref{eq:rh0Tdualsugra} produces precisely the T dual of the bosonic background of \cite{Kawaguchi:2014fca}, however the linear term in the dilaton survives the limit, and hence this model also does not correspond to supergravity, though again, it should be scale invariant.\foot{Independent of our results, the supergravity solution of \cite{Kawaguchi:2014fca} cannot correspond to the deformed sigma model as it has a Ramond-Ramond (R-R) three form that breaks the $\mathrm{SO}(6)$ invariance of the sphere that is preserved by the Yang-Baxter deformation based on $r^j_-$. This is related the direct involvement of the sphere in the duality chain of \cite{Matsumoto:2014ubv}.}

Interestingly, in \cite{vanTongeren:2015uha} it was observed that the bosonic model based on $r^j_+$ is equal to the one based on $r^j_-$ up to the sign of the deformation parameter, a total derivative in the $B$ field, and an exchange of light cone directions, but that the fermions had to differ as the deformations preserved different amounts of supersymmetry: 16 of the 32 real supercharges of $\mathfrak{psu}(2,2|4)$ for $r^j_-$ \cite{vanTongeren:2015soa} and none for $r^j_+$ \cite{vanTongeren:2015uha}. In line with this, the metric and B field of eqs. \eqref{eq:rh0Tdualsugra} are invariant under an exchange of $\hat{x}^+$ and $\hat{x}^-$ combined with a sign change on $\alpha$, while $e^\Phi F_5$ and the linear term in the dilaton single out a particular light cone direction. The fact that $r^j_-$ preserves half the supersymmetries of the $\ads$ sigma model suggests that the associated background should be a (half-maximally) supersymmetric solution of the modified supergravity equations \cite{Wulff:2016tju}.

The different scaling of $\alpha$ required to arrive at the abelian deformation based on $r^a_1$ ($\alpha \sim e^{-2\beta}$ as oppose to $\alpha \sim e^{-\beta}$ for the jordanian cases) effectively removes the linear term in the T dual dilaton so that again this does result in a solution of supergravity, in line with its interpretation as a TsT transformation \cite{vanTongeren:2015soa}. Finally, for $r^a_\pm$ -- the null version of the $r$ matrix for the HIMR background -- we start from a solution of supergravity and boost it in precisely the way that gives the gravity dual of lightlike noncommutative SYM \cite{Aharony:2000gz,Alishahiha:2000pu}.

\subsection{Overview}\label{ssec:overview}

In the above we considered a series of inequivalent singular noncompact transformations on $r^{ns}_0$. These gave the jordanian $r$ matrices
\begin{equation}
\begin{aligned}
r^j_0 = & D \wedge p^0 + M^{0}{}_{\mu} \wedge p^\mu - M^{1}{}_2\wedge p^2 - M^{1}{}_3 \wedge p^3,\\
r^j_+ = & (D + M^{-}{}_-)\wedge p^- + 2 M^{-}{}_2\wedge p^2 + 2 M^{-}{}_3\wedge p^3,\\
r^j_- = & (D + M^{+}{}_+)\wedge p^+,
\end{aligned}
\end{equation}
and the abelian $r$ matrices
\begin{equation}
\begin{aligned}
r^a_0 = & p^1 \wedge p^2,\\
r^a_- = & p^+ \wedge p^2,\\
r^a_1 = & M^{3-} \wedge p^-.
\end{aligned}
\end{equation}
These $r$ matrices preserve different amounts of symmetry: $r^j_0$ obtained by a single boost from $r^{ns}_0$ preserves a three dimensional algebra, while $r^j_+$ and $r^j_-$ -- obtained by a further boost -- preserve five dimensional algebras. For the abelian $r$ matrices we have a six dimensional algebra for $r^a_0$, and seven dimensional ones for $r^a_-$ and $r^a_1$ obtained by a second boost. The set of equivalence classes of these $r$ matrices under finite $\mathfrak{su}(2,2)$ transformations is closed under further (singular) boosts. In other words these are all the homogeneous $r$ matrices we can obtain in this way from $r^{ns}_0$.\foot{As mentioned earlier, we consider $r$ matrices with $p \leftrightarrow k$ and $D \leftrightarrow -D$ equivalent for present purposes, as this is an automorphism of the conformal algebra.}

We could, however, also start from the other non-split and split $r$ matrices, $r^{ns}_1$, $r^{ns}_2$ or $r^s$. As we discuss in more detail in appendix \ref{app:generalboosts}, the new jordanian $r$ matrices this generates are
\begin{equation}
\begin{aligned}
\tilde{r}^j_0 & =  D \wedge p^1 + M^{1}{}_{\nu} \wedge p^\nu - M^{0}{}_{2}\wedge p^2 - M^{0}{}_{3} \wedge p^3\\
r^j_1 & = D \wedge p^0 + M^{0}{}_{3} \wedge p^3,\\
\tilde{r}^j_1  & = D \wedge p^1 + M^{1}{}_{3}\wedge p^3\\
\check{r}^j_1 & = D \wedge p^1 + M^{1}{}_0 \wedge p^0,\\
r^j_{M+} & =  M^{-}{}_{-} \wedge p^{-} + M^{-}{}_2 \wedge p^2,\\
r^j_{D+} & =  D \wedge p^{-} + M^{-}{}_3 \wedge p^3,
\end{aligned}
\end{equation}
on top of which we get one further abelian $r$ matrix
\begin{equation}
\tilde{r}^a_0  = p^0 \wedge p^1.
\end{equation}
The dimensions of the undeformed symmetry algebra associated to these $r$ matrices are: three for $\tilde{r}^j_0 $, $r^j_1$, $\tilde{r}^j_1$, and $\check{r}^j_1$, four for $r^j_{M+}$ and $r^j_{D+}$ obtained by a further boost, and six for $\tilde{r}^a_0$. In total we hence find nine inequivalent (extended) jordanian $r$ matrices, and four abelian ones, from the various possible inhomogeneous $r$ matrices.\foot{We can always add abelian homogeneous terms made out of Cartan generators to our inhomogeneous $r$ matrices and continue to solve the same mcYBe. Here we do not consider such terms.} The full set of associated equivalence classes of $r$ matrices is closed under further singular boosts, and hence presents all homogeneous $r$ matrices we can obtain in this way for $\mathfrak{psu}(2,2|4)$.

As the fermions of the $r^{ns}_1$, $r^{ns}_2$ and $r^s$ deformations have strictly speaking not been considered (the $r^s$ deformation is not even real beyond bosonic $\mathrm{AdS}_5$), we cannot directly make statements about the fermions of the associated homogeneous models. However, as discussed in \cite{Hoare:2016ibq}, the bosonic backgrounds for $r^{ns}_1$ and $r^{ns}_2$ are related to the $r^{ns}_0$ one by an analytic continuation that preserves reality of the full model including the fermions, making it reasonable to assume that the resulting fermions are the ones corresponding to $r^{ns}_1$ and $r^{ns}_2$. At our jordanian level, these analytic continuations persist, cf. the relation between $r$ matrices differing by a check or a tilde. They even include the jordanian ones arising from $r^s$, since there is now no issue in (trivially) extending them to the whole of $\mathfrak{psu}(2,2|4)$. Assuming that these analytic continuations indeed give the appropriate fermions, the supergravity picture found above persists. That is, no deformation based on any of the above jordanian $r$ matrices corresponds to a solution of supergravity, while all abelian ones do. Further details on these $r$ matrices, their relations under analytic continuation, and some of the associated backgrounds can be found in appendix \ref{app:generalboosts}.

Above we indicated the dimensions of the subalgebras of $\mathfrak{su}(2,2)$ preserved by the various deformations. In terms of $\mathfrak{psu}(2,2|4)$, the $\mathfrak{su}(4)$ factor is always preserved, while supersymmetry is broken completely in all of the jordanian deformations with the exception of the one based on $r^j_-$ discussed in section \ref{sec:consecutiveboosts} above.

This exercise can be repeated for lower dimensional cases such as the $\mathrm{AdS}_3$ and $\mathrm{AdS}_2$ string, which results in $r$ matrices corresponding to appropriate truncations of the above. The supergravity picture based on the corresponding solutions for the T duals of $\eta$-deformed  $\mathrm{AdS}_3 \times \mathrm{S}^3$ and $\mathrm{AdS}_2 \times \mathrm{S}^2$ strings is then the same: jordanian ones do not correspond to supergravity, while abelian ones do. It is perhaps worth pointing out that for $\mathrm{AdS}_2$ there is only one homogeneous $r$ matrix, the jordanian $r= D\wedge p^0$, which follows from the (unique) non-split one by our procedure.\foot{It can also be obtained from the split $r$ matrix.} Hence, in this case we can derive all homogeneous Yang-Baxter deformations from the non-split deformation, none of which correspond to supergravity.

Let us emphasize that for larger algebras we definitely cannot find all homogeneous $r$ matrices by boosts. Any $r$ matrix we obtain in this way satisfies $r^3 = 0$. This is an immediate consequence of relation $r^3 = \mp r$ that the non-split and split $r$ matrices satisfy.\foot{The previously mentioned abelian homogenous terms that can be added to these inhomogeneous $r$ matrices would break this relation.} However, we do not even find all $r$ matrices solving this condition. Perhaps most striking example in this regard is the extended jordanian $\mathfrak{su}(2,2)$ $r$ matrix
\begin{equation}
\label{eq:leastsymmrmatrix}
r = D \wedge p^0 + M^{0}{}_{1} \wedge p^1 + M^{0}{}_{2} \wedge p^2 - M^{1}{}_2\wedge p^2,
\end{equation}
obtained by simply dropping the $p_3$ terms in $r^j_0$. This $r$ matrix breaks the rotational symmetry in the $23$-plane preserved by $r^j_0$ and preserves only two generators of $\mathfrak{su}(2,2)$, one less than any of the inhomogeneous $r$ matrices.

\section{Concluding remarks}\label{sec:conclusions}

In this paper we considered the possible inequivalent infinite boost limits of various non-split and split $r$ matrices and the associated deformations of $\ads$, generating various homogeneous $r$ matrices of both jordanian and abelian type and their associated deformed backgrounds. This allowed us to discuss the fermions of these models based on those associated to the $\eta$-deformed string, in certain cases up to analytic continuation. The upshot of the corresponding analysis is that the jordanian deformations we obtain in this way are analogous to the $\eta$ deformation in terms of supergravity \cite{Arutyunov:2015qva,Hoare:2015wia}: T dualizing them at the classical level they correspond to solutions of supergravity. These solutions, however, have dilatons that prevent T dualizing back and hence the models themselves do not correspond to supergravity. These jordanian backgrounds are solutions of the modified supergravity equations proposed in \cite{Arutyunov:2015mqj} and as such should nevertheless be scale invariant.\footnote{Let us also recall that the jordanian deformation based on $r^j_-$ preserves half the supersymmetry of the undeformed string \cite{vanTongeren:2015soa,vanTongeren:2015uha}, suggesting it should be a supersymmetric solution of the modified equations \cite{Wulff:2016tju}.} This may also be expected as a consequence of the $\kappa$ symmetry of the models. These results give a number of counterexamples to the conjecture that Yang-Baxter deformations of the $\ads$ superstring based on solutions of the cYBe always correspond directly to superstring theories in supergravity backgrounds (thus far referred to as the ``gravity/cYBe'' correspondence in the literature).
In contrast to jordanian deformations, the abelian deformations we obtain in this way are solutions of supergravity, as expected. In addition to this we briefly investigated contraction limits of our main model, confirming previous suggestions \cite{Pachol:2015mfa,Hoare:2016ibq}. These contractions do correspond to supergravity, in contrast to the situation with the $\eta$-deformed string \cite{Arutyunov:2015qva}.\foot{Here and at some points below it may be relevant to keep in mind that the bosonic part of the contraction limit of the $\eta$ model can be completed to a solution of supergravity, giving the so-called mirror model  \cite{Arutynov:2014ota,Arutyunov:2014cra,Arutyunov:2014jfa}. The mirror model is an integrable model itself, and is closely related to the direct contraction of the full $\eta$ model \cite{Arutyunov:2015qva}.}

There are a number of open questions associated to these deformations. First, our approach is by construction limited to the set of homogeneous deformations we discussed. This does not include all jordanian deformations, and it would be interesting to investigate whether these other deformations have distinguishing features. It would also be good to investigate other deformed sigma models in detail -- e.g. the one associated to the $r$ matrix \eqref{eq:leastsymmrmatrix} -- perhaps by more direct methods such as a full supercoset construction. Second, related to this, perhaps there are other ways to relate inhomogeneous deformations to homogeneous deformations that give new cases.\foot{We emphasize that our method and the classification in section \ref{ssec:overview} appear to exhaust all singular group transformations.} Third, before it was clear that the supergravity solution of \cite{Kawaguchi:2014fca,Matsumoto:2014ubv} only agrees with the jordanian deformed model at the bosonic level, it was found that this supergravity solution can be obtained from a deformation of a D3 brane stack in a low energy limit, giving it an interpretation as the gravity dual of a noncommutative version of SYM \cite{vanTongeren:2015soa}. In fact, the corresponding noncommutative structure is compatible with the Drinfeld twisted structure of the jordanian model \cite{vanTongeren:2015soa}. It would be interesting to clarify this point further, to see what distinguishes this gravity dual from the hypothetical Drinfeld twisted model, and whether some aspects of integrability may nevertheless present themselves in this AdS/CFT setting. Fourth, since the jordanian deformations we consider are similar to the $\eta$ deformation in the sense that they satisfy the modified supergravity equations of \cite{Arutyunov:2015mqj}, many of the open questions there become relevant here as well. For instance: can these scale invariant but (presumably) not Weyl invariant models still be used to define critical string theories? what is the interplay between the modified supergravity equations and the $\kappa$ symmetry of these models? what do we find if we analyze the Weyl invariance conditions directly from the sigma model action? Inspired by the Poisson-Lie duality \cite{Klimcik:1995ux,Klimcik:1995jn,Vicedo:2015pna,Hoare:2015gda,Sfetsos:2015nya,Klimcik:2015gba} between the $\eta$ type models and the $\lambda$ models of \cite{Sfetsos:2013wia,Hollowood:2014qma,Demulder:2015lva} we might also ask whether homogeneous deformations exist in the spirit of the $\lambda$ model. One approach might be to try to take the infinite boosts directly in the $\lambda$ model. However, this does not appear to work in a simple manner. Another approach would be to directly construct the Poisson-Lie dual \cite{Klimcik:1995ux,Klimcik:1995jn} of the backgrounds constructed from homogeneous $r$ matrices. Finally, going a bit beyond present considerations, while the jordanian models are integrable by construction, they invariably break the isometries required to fix the standard BMN light cone gauge central to the exact S matrix approach to the quantum string sigma model \cite{Arutyunov:2009ga}. In other words, the effect of these deformations at the quantum level is mysterious, in contrast to e.g. the $\beta$ deformation \cite{vanTongeren:2013gva}. It would also be interesting to investigate deformations associated to homogeneous $r$ matrices containing fermionic generators.

\section*{Acknowledgments}

We would like to thank G. Arutyunov, S. Frolov, and A. A. Tseytlin for comments on the draft. The work of B.H. is partially supported by grant no. 615203 from the European Research Council under the FP7. S.T. is supported by L.T. The work of S.T. is supported by the Einstein Foundation Berlin in the framework of the research project "Gravitation and High Energy Physics" and acknowledges further support from the People Programme (Marie Curie Actions) of the European Union's Seventh Framework Programme FP7/2007-2013/ under REA Grant Agreement No 317089.

\appendix

\section{General boosts of inhomogeneous $r$ matrices}

\label{app:generalboosts}

In this appendix we consider general boosts of the four inhomogeneous $r$ matrices introduced in section \ref{sec:exampleintro}.

\subsection{Non-split}

As mentioned in the main text, we can take $b_{01}$ and $b_{03}$ to represent the inequivalent classes of boosts we can apply to our non-split $r$ matrices. Including $r^{ns}_0$ for completeness, acting with $b_{01}$ on the three non-split $r$ matrices and scaling the deformation parameter as indicated gives
\begin{equation}
\begin{aligned}
\lim_{\beta \rightarrow \infty}2 \alpha e^{-\beta} \mbox{Ad}_{b_{01}}(r^{ns}_0) &  = \alpha r^j_0,\\
\lim_{\beta \rightarrow \infty}2 \alpha e^{-\beta} \mbox{Ad}_{b_{01}}(r^{ns}_1) &  = -\alpha \tilde{r}^j_0,\\
\lim_{\beta \rightarrow \infty}2 \alpha e^{-\beta} \mbox{Ad}_{b_{01}}(r^{ns}_2) &  = -\alpha \check{r}^j_1,
\end{aligned}
\end{equation}
where
\begin{equation}
\tilde{r}^j_0 =  D \wedge p^1 + M^{1}{}_{\nu} \wedge p^\nu - M^{0}{}_{2}\wedge p^2 - M^{0}{}_{3} \wedge p^3,
\end{equation}
and
\begin{equation}
\check{r}^j_1 = D \wedge p^1 + M^{1}{}_0 \wedge p^0,
\end{equation}
which are further $r$ matrices of extended jordanian type. Note that $\tilde{r}^j_0$ is related to $r^j_0$ of eqn. \eqref{eq:rh0} by the analytic continuation $0\leftrightarrow1$.

Boosting with $b_{03}$ instead we get
\begin{equation}
\begin{aligned}
\lim_{\beta \rightarrow \infty}4 \alpha e^{-2\beta} \mbox{Ad}_{b_{03}}(r^{ns}_0) &  = \alpha r^a_0,\\
\lim_{\beta \rightarrow \infty}2 \alpha e^{-\beta} \mbox{Ad}_{b_{03}}(r^{ns}_1) &  = \alpha r^j_1,\\
\lim_{\beta \rightarrow \infty}4 \alpha e^{-2\beta} \mbox{Ad}_{b_{03}}(r^{ns}_2) &  = -\alpha r^a_0,
\end{aligned}
\end{equation}
where
\begin{equation}
r^j_1  = D \wedge p^0 + M^{0}{}_{3} \wedge p^3,
\end{equation}
which is related to $\check{r}^j_1$ by analytically continuing $0\leftrightarrow3$ and rotating $3\rightarrow1$. Let us briefly discuss the backgrounds for $\tilde{r}^j_0$, $r^j_1$ and $\check{r}^j_1$; the backgrounds for $r^j_0$ and $r^a_0$ are discussed extensively in the main text.

\subsubsection*{The background for $\tilde{r}^j_0$}

The background associated to  $\tilde{r}^j_0$ is
\begin{equation}
\begin{aligned}
\label{eq:hombackground0t}
ds^2_1 & = \frac{dz^2+dx^2}{z^2 + \alpha^2} + \frac{-dt^2 + dr^2}{z^2 - \alpha^2 r^2/z^2} +\frac{r^2 d\theta^2}{z^2},\\
B_1 & = \frac{\alpha}{z}\frac{1}{z^2+\alpha^2} dx \wedge dz - \frac{\alpha r}{z^4-\alpha^2 r^2} dt \wedge dr,
\end{aligned}
\end{equation}
which is related to the background for $r^{ns}_1$ as the one for $r^j_0$ is to the $r^{ns}_0$ one. The limit discussed in the main text gives the above background from the corresponding non-split one that can be found in \cite{Hoare:2016ibq}. The coset construction picture in terms of $g_1$ of \cite{Hoare:2016ibq} is that the constant $\pi/2$ shift in $t$ that we need for the (same) $\rho$ part of the group element, precisely rotates $m^{35}$ to $m^{03}$, so that the (different) $x$ part of the group element boosts as before, in line with the limit. Alternatively we can work with the nondiagonal metric for the coset model based on $(r^{ns}_1,g_0)$ that can be found in \cite{Delduc:2014kha} and take the limit discussed for $g_0$ in the main text.

As discussed in the main text, the fermions for the $r^{ns}_1$ deformation have not been directly investigated, however we may assume they can be obtained by analytic continuation from the $r^{ns}_0$ ones. At the current jordanian level the statement is simply that our $\tilde{r}^j_0$ background is related to the $r^j_0$ one of eqn. \eqref{eq:hombackground0} by the analytic continuation $x\leftrightarrow i t$ and $\alpha \rightarrow i \alpha$, cf. the relation between the $r$ matrices. Since this analytic continuation preserves reality of the full (T dual) background for $r^j_1$, and by definition preserves integrability, the resulting background has a good chance to correspond to the fermions of the deformed sigma model, which again would not be compatible with supergravity.

\subsubsection*{The backgrounds for $r^j_1$ and $\check{r}^j_1$}

To find the backgrounds associated to $r^j_1$ and $\check{r}^j_1$, we can just apply the limit discussed in the main text to the non-split sigma models based on $(r^{ns}_1,g_0)$ and $(r^{ns}_2,g_0)$.\foot{This is an alternative to working out the relation between the coset parametrizations $g_1$ and $g_2$ of \cite{Hoare:2016ibq}, and $g^h$.} Doing so gives
\begin{equation}
\begin{aligned}
ds_{3}^2 & =  \frac{dx^2+r^2 d\theta^2}{z^2}+ \frac{z^2}{z^4 - \alpha^2(r^2 + z^2)}\left(-dt^2 + dr^2 + dz^2 - \frac{\alpha^2}{z^2}(dr - \frac{r}{z} dz)^2 \right),\\
B_{3} & = \alpha \frac{r dr \wedge dt + z dz \wedge dt}{z^4 - \alpha^2(r^2 +z^2)},
\end{aligned}
\end{equation}
and
\begin{equation}
\begin{aligned}
ds_2^2 & =  \frac{-dt^2+r^2 d\theta^2}{z^2}+ \frac{z^2}{z^4 + \alpha^2(r^2 + z^2)}\left(dx^2 + dr^2 + dz^2 + \frac{\alpha^2}{z^2}(dr - \frac{r}{z} dz)^2 \right),\\
B_2 & = \alpha \frac{r dr \wedge dx + z dz \wedge dx}{z^4 + \alpha^2(r^2 +z^2)},
\end{aligned}
\end{equation}
respectively, which are also precisely the backgrounds of the homogeneous sigma models based on $(\check{r}^j_1,g^h)$ and $(r^j_1,\tilde{g}^h)$ respectively, where $\tilde{g}^h$ means $g^h$ written in physical generators with indices $1$ and $3$ exchanged.\foot{This in line with the above mentioned exchange of $\psi_1$ and $\psi_2$ when boosting by $b_{03}$ as opposed to $b_{01}$. Alternatively, instead of using $r^j_1  = D \wedge p^0 + M^{0}{}_{3} \wedge p^3$ we could use the physically equivalent $D \wedge p^0 + M^{0}{}_{1} \wedge p^1$ and work with $g^h$.} As expected from their $r$ matrices, these backgrounds are also related by the analytic continuation $x \leftrightarrow i t$, $\alpha \rightarrow i \alpha$.

In order to discuss the fermions for these deformed models, we would again need knowledge of the fermions of the $r^{ns}_1$ and $r^{ns}_2$ deformations. As for the $\tilde{r}^j_0$ background we may assume that these are obtained by analytic continuation, in which case the linear term in the dilaton of the T dual backgrounds again survives and the fermions of these jordanian deformed models again would not correspond to supergravity.

\subsection{Split}

The split $r$ matrix \eqref{eq:splitrmatrix} is only invariant under a permutation of indices $3$ and $4$, which would leave six inequivalent boosts: $b_{01}$, $b_{02}$, $b_{03}$, $b_{51}$, $b_{52}$, and $b_{53}$. It is however invariant under the adjoint action of $b_{02}$ and $b_{51}$. This leaves
\begin{equation}
\begin{aligned}
\lim_{\beta \rightarrow \infty}2 \alpha e^{-\beta} \mbox{Ad}_{b_{01}}(r^{s}) &  = \alpha r^j_0,\\
\lim_{\beta \rightarrow \infty}2 \alpha e^{-\beta} \mbox{Ad}_{b_{03}}(r^{s}) &  = -\alpha \tilde{r}^j_1,\\
\lim_{\beta \rightarrow \infty}2 \alpha e^{-\beta} \mbox{Ad}_{b_{52}}(r^{s}) &  = -\alpha \tilde{r}^j_0,\\
\lim_{\beta \rightarrow \infty}4 \alpha e^{-2\beta} \mbox{Ad}_{b_{53}}(r^{s}) &  = \alpha \tilde{r}^a_0,
\end{aligned}
\end{equation}
where the new $r$ matrices we find are
\begin{equation}
\tilde{r}^j_1  = D \wedge p^1 + M^{1}{}_{3}\wedge p^3,
\end{equation}
and
\begin{equation}
\tilde{r}^a_0  = p^0 \wedge p^1,
\end{equation}
which are the analytic continuations $0\leftrightarrow1$ of $r^j_1$ and $2 \leftrightarrow 0$ of $r^a_0$ respectively. The associated bosonic backgrounds are related by the same analytic continuations. Assuming the fermions can indeed be obtained by the same analytic continuation, this jordanian deformation also would not correspond to supergravity, while the abelian one does.

\subsection{Further Lorentz boosts}

We can now Lorentz boost all above $r$ matrices. In terms of new inequivalent $r$ matrices this yields
\begin{equation}
r^j_{M+} =  M^{-}{}_{-} \wedge p^{-} + M^{-}{}_2 \wedge p^2,
\end{equation}
and
\begin{equation}
r^j_{D+} =  D \wedge p^{-} + M^{-}{}_3 \wedge p^3,
\end{equation}
which are obtained by acting with $B_{03}$ on $\tilde{r}^j_0$ and $B_{01}$ on $\tilde{r}^j_{1}$ respectively. The scaling of the deformation parameter here is identical to the other jordanian cases, and hence the associated models do not correspond to supergravity, provided the analytic continuations described above give the correct fermions for the $\tilde{r}^j_0$ and $\tilde{r}^j_1$ models.

\section{Contractions of the $r^j_0$ deformed model}

\label{app:contractions}

As $\mathfrak{so}(2,d-1)$ can be contracted to the $d$-dimensional Poincar\'e algebra, the corresponding quantum group can be contracted to what is known as the $\kappa$-Poincar\'e algebra \cite{Lukierski:1991pn}. A similar thing can be done for the quantum group symmetry of the $\eta$-deformed $\ads$ superstring \cite{Pachol:2015mfa}. In the undeformed case this contraction geometrically amounts to going from $\ads$ to flat space, and hence the $\eta$-deformed model contracts to what can be though of as a deformation of the flat space string. In our previous paper \cite{Hoare:2016ibq} we discussed contractions of the various inhomogeneous deformations of $\mathrm{AdS}_5$, and since our main example of section \ref{sec:exampleintro} is so closely related to the standard non-split and split deformations -- $r^j_0 = r^{ns}_0 + r^s$ -- we would like to briefly address the corresponding contractions here.

As discussed in more detail in \cite{Pachol:2015mfa,Hoare:2016ibq} we will consider the generators $m^{ij}$ and select one index $\hat{\imath}$ and scale the generators containing that index to infinity
\begin{equation}\label{mscale}
m^{j\hat{\imath}} \rightarrow \mathrm{R} m^{j\hat{\imath}} \ , \qquad \mathrm{R}  \rightarrow \infty \ .
\end{equation}
Starting from undeformed $\mathfrak{su}(2,2) \simeq \mathfrak{so}(2,4)$ this gives $\mathfrak{iso}(1,4)$ if $\hat{\imath}$ is a timelike index ($0$ or $5$), or $\mathfrak{iso}(2,3)$ if it is spacelike ($1,\ldots,4$). These contractions give nontrivial results in a $q$ deformed algebra if we additionally scale $q$ as $\log q \to - \mathrm{R} ^{-1} \log q$. If we contract with a timelike index this gives $\mathcal{U}_{\kappa}(\mathfrak{iso}(1,4))$, a deformation of the 5-d Poincar\'e algebra, while with a spacelike index we get $\mathcal{U}_{\kappa}(\mathfrak{iso}(2,3))$. Both are $\kappa$-Poincar\'e algebras, see e.g. \cite{Borowiec:2013lca} for a unified discussion of $\kappa$-Poincar\'e algebras for any dimension and signature.

Referring to \cite{Pachol:2015mfa,Hoare:2016ibq} for details, we simply note here that we can translate these contractions to the background by looking at the coset representative to determine the scaling of the coordinates required to keep a finite action, up to a rescaling of the effective string tension by $\mathrm{R}^2$ that is already present in the undeformed flat space case. To keep the contraction manifestly compatible with the structure of the action \eqref{eq:defaction}, the natural contraction index is the coset direction $5$  ($0$ for the physical generators). Looking at $g^h$, after the coordinate transformation $\tilde{t} = -t+z$ we can sensibly implement this contraction by rescaling $\tilde{t}$ and $z$ as $\tilde{t}\rightarrow \tilde{t}/\mathrm{R} $ and $z \rightarrow \mathrm{R}  z$, and considering the limit $\mathrm{R}  \rightarrow \infty$ with $\alpha = \kappa^{-1} \mathrm{R} $. This gives
\begin{equation}
ds^2 = \frac{ 2 dz d\tilde{t}}{z^2-\kappa^{-2}} + \frac{dx^2 + dr^2 + r^2 d\theta^2}{z^2},
\end{equation}
while the $B$ field becomes a total derivative. At $\kappa^{-1}=0$ this is flat space, albeit written in curious coordinates. In \cite{Pachol:2015mfa} it was suggested that this deformation and its contraction should exist, that the associated $r$ matrix should be the one of the ``null'' $\kappa$-Poincar\'e algebra, and that the background should correspond to a five dimensional plane wave.\footnote{The two dimensional analogue of the above background obtained by truncating to constant $x$, $r$, and $\theta$ was explicitly given in \cite{Pachol:2015mfa}, in this low dimensional setting the plane wave becomes just flat space in suggestive coordinates however.} Moreover, the contemporary results of \cite{Borowiec:2015wua} showed that using the four dimensional null $\kappa$-Poincar\'e $r$ matrix to deform four dimensional flat space directly, indeed gives a plane wave. To make contact with this result we can define $\tilde{t} = -\frac{2 x^- x^+ + \tilde{r}^2}{2x^-}$, $z = 1/x^-$, $x\rightarrow \frac{\tilde{r} \sin \tilde{\theta} \cos \tilde{\phi}}{x^-}$, $r= \frac{\tilde{r}\sqrt{\sin^2 \tilde{\theta} \sin^2 \tilde{\phi} + \cos^2 \tilde{\theta}}}{x^-}$, and $\theta = \arctan (\sin \tilde{\phi} \tan \tilde{\theta})$, which gives
\begin{equation}
ds^2 = \frac{dx^+ dx^- + \kappa^{-2} (2 x^- \tilde{r} dx^- d\tilde{r} -(dx^-)^2)}{1-\kappa^{-2} (x^-)^2}+ d\tilde{r}^2 + \tilde{r}^2 d\tilde{\theta}^2 + \tilde{r}^2 \sin^2{\tilde{\theta}} d\tilde{\phi}^2.
\end{equation}
This is precisely the five dimensional analogue of the plane wave presented in section 4.3 of \cite{Borowiec:2015wua}, and reduces manifestly to flat space in ``light-cone-spherical'' coordinates at $\kappa^{-1}=0$.

Though less obvious, there are further possible contractions of a particular type discussed in \cite{Hoare:2016ibq}. In the current context the corresponding contraction is over index $4$ ($3$ for the physical generators). This corresponds to rescaling only $\theta$ by $\mathrm{R}$ and gives
\begin{equation}
ds^2 = \kappa^2 (-dz^2 + dt^2 + \frac{z^2}{r^2}(dx^2 + dr^2)) + \frac{r^2}{z^2} d\theta^2,
\end{equation}
up to an infinite total derivative $B$ field. Note that this limit has taken us out of the physical range of the coordinates in the original model, that correspondingly $z$ has become timelike, and that the deformation and contraction limit do not commute here since $\kappa^{-1} = 0$ is a singular point of the metric.

In \cite{Pachol:2015mfa,Hoare:2016ibq} it was observed that the corresponding contractions of inhomogeneous deformations are T dual to $\mathrm{dS}_5$ and $\mathrm{AdS}_5$, depending on whether one starts with a non-split or split deformation respectively, and the T duality involves the isometry coordinate involved in the contraction. In the first contraction above, because $\tilde{t}$ is a null direction it is not clear whether there is something that could take on the role of this T duality. In the second case, however, we can perfectly T dualize in $\theta$, which as we might have guessed gives the ``midpoint'' between $\mathrm{dS}_5$ and $\mathrm{AdS}_5$: flat space.

Implementing these contraction limits in the singular boost limit of the full background of \cite{Arutyunov:2015qva} we find that the R-R fluxes vanish. The first background can now be supported by the dilaton $\Phi = \tfrac{3}{2}\log(1-\tfrac{1}{\kappa^2 z^2}) + c \arctanh{\kappa z} + \Phi_0 $, where $c$ and $\Phi_0$ are constants. The contraction limit of the jordanian limit of $X$ of \cite{Arutyunov:2015mqj} gives this dilaton with $c=4$.\foot{The one form $I$ of \cite{Arutyunov:2015mqj} remains finite in this limit, but as all R-R forms vanish this background can nevertheless be a solution of supergravity. The one form $I$ is exact and in fact integrates to $\arctanh{\kappa z}$, cf. the dilaton.} In the second case we find $\Phi = \log \tfrac{r}{z} +  c \tfrac{z}{r} +\Phi_0$, where $c=0$ corresponds to the limit of the results of \cite{Arutyunov:2015mqj}, giving nothing but the dilaton obtained by directly T dualizing flat space with a constant (bounded) dilaton, in $\hat{\theta}$. In both cases the freedom in the dilaton parametrised by $c$ corresponds to a null direction of the metric. It would be interesting to understand this freedom from the perspective of the integrable Yang-Baxter sigma models; is there a principle that forces us to pick the dilatons as those coming from the $\eta$ model? Let us also note that the fact that these contractions give supergravity solutions is in contrast to the standard contraction limit of the $\eta$-deformed string for which this is not the case \cite{Arutyunov:2015qva}, though we should mention that the bosonic part of the $\eta$-deformed model can be completed to a solution of supergravity in a way that is compatible with integrability, corresponding to the so-called mirror model \cite{Arutynov:2014ota,Arutyunov:2014cra,Arutyunov:2014jfa}. Some of the other deformations considered in this paper can be contracted in similar fashion.


\bibliographystyle{jhep}

\bibliography{stijnsbibfile}


\end{document}